\documentclass[aps,prl,twocolumn,superscriptaddress]{revtex4}
\usepackage{graphicx}
\usepackage{dcolumn}

\date{March, 31}                                                                                          
\begin{document}
                                                                                         
\title{\large Electroexcitation of the Roper resonance for 
              $1.7< Q^2< 4.5~$GeV$^2$ 
              in $\vec{e}p\rightarrow en\pi^+$}
                                                                                         
\newcommand*{\JLAB}{Thomas Jefferson National Accelerator Facility, 
Newport News, Virginia 23606}
\affiliation{\JLAB}
\newcommand*{\YEREVAN}{Yerevan Physics Institute, 375036 Yerevan, 
Armenia}
\affiliation{\YEREVAN}
\newcommand*{\SCAROLINA}{University of South Carolina, Columbia, South 
Carolina 29208}
\affiliation{\SCAROLINA}
\newcommand*{\KYUNGPOOK}{Kyungpook National University, Daegu 702-701, 
Republic of Korea}
\affiliation{\KYUNGPOOK}
\newcommand*{\ANL}{Argonne National Laboratory, Argonne,
Illinois 60439}
\affiliation{\ANL}
\newcommand*{\ASU}{Arizona State University, Tempe, Arizona 85287-1504}
\affiliation{\ASU}
\newcommand*{\UCLA}{University of California at Los Angeles, Los 
Angeles, California  90095-1547}
\affiliation{\UCLA}
\newcommand*{\CSU}{California State University, Dominguez Hills, Carson, 
California 90747}
\affiliation{\CSU}
\newcommand*{\CMU}{Carnegie Mellon University, Pittsburgh, Pennsylvania 
15213}
\affiliation{\CMU}
\newcommand*{\CUA}{Catholic University of America, Washington, D.C. 
20064}
\affiliation{\CUA}
\newcommand*{\SACLAY}{CEA-Saclay, Service de Physique Nucl\'eaire, 91191 
Gif-sur-Yvette, France}
\affiliation{\SACLAY}
\newcommand*{\CNU}{Christopher Newport University, Newport News, 
Virginia 23606}
\affiliation{\CNU}
\newcommand*{\UCONN}{University of Connecticut, Storrs, Connecticut 
06269}
\affiliation{\UCONN}
\newcommand*{\Chile}{Universidad T\'ecnica Federico Santa Mar\'ia, 
Casilla 110-V, Valpara\'iso, Chile}
\affiliation{\Chile}
\newcommand*{\ECOSSEE}{Edinburgh University, Edinburgh EH9 3JZ, United 
Kingdom}
\affiliation{\ECOSSEE}
\newcommand*{\EMMY}{Emmy-Noether Foundation, Germany}
\affiliation{\EMMY}
\newcommand*{\FU}{Fairfield University, Fairfield, Connecticut 06824}
\affiliation{\FU}
\newcommand*{\FIU}{Florida International University, Miami, Florida 
33199}
\affiliation{\FIU}
\newcommand*{\FSU}{Florida State University, Tallahassee, Florida 32306}
\affiliation{\FSU}
\newcommand*{\GWU}{The George Washington University, Washington, D.C. 
20052}
\affiliation{\GWU}
\newcommand*{\ECOSSEG}{University of Glasgow, Glasgow G12 8QQ, United 
Kingdom}
\affiliation{\ECOSSEG}
\newcommand*{\ISU}{Idaho State University, Pocatello, Idaho 83209}
\affiliation{\ISU}
\newcommand*{\INFNFR}{INFN, Laboratori Nazionali di Frascati, 00044 
Frascati, Italy}
\affiliation{\INFNFR}
\newcommand*{\INFNGE}{INFN, Sezione di Genova, 16146 Genova, Italy}
\affiliation{\INFNGE}
\newcommand*{\ORSAY}{Institut de Physique Nucleaire ORSAY, Orsay, 
France}
\affiliation{\ORSAY}
\newcommand*{\ITEP}{Institute of Theoretical and Experimental Physics, 
Moscow, 117259, Russia}
\affiliation{\ITEP}
\newcommand*{\JMU}{James Madison University, Harrisonburg, Virginia 
22807}
\affiliation{\JMU}
\newcommand*{\MIT}{Massachusetts Institute of Technology, Cambridge, 
Massachusetts  02139-4307}
\affiliation{\MIT}
\newcommand*{\UMASS}{University of Massachusetts, Amherst, Massachusetts  
01003}
\affiliation{\UMASS}
\newcommand*{\MOSCOW}{Moscow State University, General Nuclear Physics 
Institute, 119899 Moscow, Russia}
\affiliation{\MOSCOW}
\newcommand*{\UNH}{University of New Hampshire, Durham, New Hampshire 
03824-3568}
\affiliation{\UNH}
\newcommand*{\NSU}{Norfolk State University, Norfolk, Virginia 23504}
\affiliation{\NSU}
\newcommand*{\OHIOU}{Ohio University, Athens, Ohio  45701}
\affiliation{\OHIOU}
\newcommand*{\ODU}{Old Dominion University, Norfolk, Virginia 23529}
\affiliation{\ODU}
\newcommand*{\PITT}{University of Pittsburgh, Pittsburgh, Pennsylvania 
15260}
\affiliation{\PITT}
\newcommand*{\RPI}{Rensselaer Polytechnic Institute, Troy, New York 
12180-3590}
\affiliation{\RPI}
\newcommand*{\RICE}{Rice University, Houston, Texas 77005-1892}
\affiliation{\RICE}
\newcommand*{\URICH}{University of Richmond, Richmond, Virginia 23173}
\affiliation{\URICH}
\newcommand*{\TRIUMF}{TRIUMF, 4004, Wesbrook Mall, Vancouver, BC, V6T 
2A3, Canada}
\affiliation{\TRIUMF}
\newcommand*{\UNIONC}{Union College, Schenectady, New York 12308}
\affiliation{\UNIONC}
\newcommand*{\VT}{Virginia Polytechnic Institute and State University, 
Blacksburg, Virginia   24061-0435}
\affiliation{\VT}
\newcommand*{\VIRGINIA}{University of Virginia, Charlottesville, 
Virginia 22901}
\affiliation{\VIRGINIA}
\newcommand*{\WM}{College of William and Mary, Williamsburg, Virginia 
23187-8795}
\affiliation{\WM}
\newcommand*{\deceased}{Deceased}

\author{I.G.~Aznauryan}
     \affiliation{\JLAB}
     \affiliation{\YEREVAN}
\author{V.D.~Burkert}
     \affiliation{\JLAB}
\author{W.~Kim}
     \affiliation{\KYUNGPOOK}
\author {K.~Park} 
\affiliation{\SCAROLINA}
\affiliation{\KYUNGPOOK}
\author {G.~Adams} 
\affiliation{\RPI}
\author {M.J.~Amaryan} 
\affiliation{\ODU}
\author {P.~Ambrozewicz} 
\affiliation{\FIU}
\author {M.~Anghinolfi} 
\affiliation{\INFNGE}
\author {G.~Asryan} 
\affiliation{\YEREVAN}
\author {H.~Avakian} 
\affiliation{\JLAB}
% ...
 
\collaboration{The CLAS Collaboration}
     \noaffiliation

\begin{abstract} 
{The helicity amplitudes of the electroexcitation
of the Roper resonance are extracted
for $1.7< Q^2< 4.5~$GeV$^2$ from recent high precision
JLab-CLAS  cross section and longitudinally polarized
beam asymmetry data for $\pi^+$ electroproduction on protons
at $W=1.15-1.69~$GeV.
The analysis is made using two approaches, dispersion relations
and a unitary isobar model, which give consistent results.
It is found that the transverse helicity amplitude $A_{1/2}$
for the $\gamma^* p\rightarrow ~$N(1440)P$_{11}$ transition,
which is large and negative at $Q^2=0$,
becomes large and positive at $Q^2\simeq 2~$GeV$^2$,
and then drops slowly with $Q^2$. 
The longitudinal helicity amplitude $S_{1/2}$, which was
previously found from CLAS $\vec{e}p\rightarrow ep\pi^0,en\pi^+$
data to be large and positive at
$Q^2=0.4,~0.65~$GeV$^2$, drops
with  $Q^2$. Available model predictions for
$\gamma^* p\rightarrow ~$N(1440)P$_{11}$ allow us 
to conclude that these results
provide strong evidence in favor
of N(1440)P$_{11}$ as a first radial excitation
of the 3$q$ ground state. The results of the present paper 
also confirm
the conclusion of our previous analysis for $Q^2< 1~$GeV$^2$
that the presentation of N(1440)P$_{11}$ as a q$^3$G
hybrid state is ruled out.} 
\end{abstract}

\maketitle

The excitation of nucleon resonances
in electromagnetic interactions has long been recognized
as a sensitive source of information on the long-
and short-range structure of the nucleon and its excited
states in the domain of quark confinement. 
Constituent quark models (CQM) have been developed
that relate electromagnetic resonance transition form factors
to fundamental quantities, such as the quark confining potential.
While this relationship is more direct for heavy quarks,
even in the light quark sector such connections
exist and may be probed by measuring transition
form factors over a large range in photon
virtuality $Q^2$, which 
defines the space-time resolution of the probe.

The so-called Roper resonance, or N(1440)P$_{11}$,
is the lowest excited state of the nucleon.
In the CQM, the simplest and most natural
assumption is that this is the first
radial excitation of the 3$q$ ground state. However,
calculations within the nonrelativistic CQM
fail to reproduce even the sign of the transition
photo-coupling amplitude \cite{Capstick1}.
Moreover, the mass of the state is more than 100 MeV lower
than what is predicted in the CQM with gluon exchange
interaction \cite{Capstick2,Richard}.
More recent models that include also Goldstone boson 
exchange between quarks gave better agreement with the mass 
\cite{Glozman}.  
To deal with shortcomings of the quark model,
alternative descriptions of 
N(1440)P$_{11}$ were
developed, where this resonance is treated respectively 
as: a hybrid q$^3$G state where the three
quarks are bound together with a gluon   \cite{Li1,Li2},
a quark core dressed by a meson cloud
\cite{Cano1,Cano2}, and a dynamically generated $\pi N$ 
resonance \cite{Krewald}; other models 
include 3$q-q\bar{q}$ components,
in particular a strong $\sigma N$ component (see Ref. 
\cite{Dillig}
and references therein).
Discrimination between these descriptions 
of the Roper resonance can provide deep insight
into the underlying basic symmetries and 
the structure of quark confinement.

The $Q^2$ dependence of the electromagnetic transition
form factors is highly sensitive to different
descriptions of the Roper state.
However, until recently, the data base used to extract
these form factors was almost exclusively based on $\pi^0$
production, and very limited in kinematical coverage.
Also, the $\pi^0 p$ final state is dominated by the nearby  
isospin $\frac{3}{2}$ $\Delta(1232)$P$_{33}$ resonance, 
whereas the isospin $\frac{1}{2}$ Roper state
couples more strongly to the $\pi^+ n$ channel.
The CLAS Collaboration has now published a large body
of precise differential cross sections and polarized beam asymmetries
for the process $\vec{e}p\rightarrow en\pi^+$
in the range of invariant hadronic mass
$W=1.15-1.69~$GeV and photon virtuality
$Q^2=1.7-4.5~$GeV$^2$, with full azimuthal and polar
angle coverage \cite{Park}. 
In this Letter we report the results on the electroexcitation
of the Roper resonance  extracted from this large data set.

The approaches we used to analyze the data are 
fixed-$t$
dispersion relations (DR) and a unitary isobar model (UIM). 
They
were successfully employed in Refs. \cite {Azn0,Azn04,Azn065} 
for analyses of pion-photoproduction 
and low-$Q^2$-electroproduction data.  

The imaginary parts of the amplitudes in the 
DR and UIM approaches
are determined mainly by $s$-channel resonance contributions
that we parameterize in the usual Breit-Wigner form
with energy-dependent widths. We also take  
into account inelastic channels in the form 
proposed in Ref. \cite{Drechsel}. An 
exception
was made  for
the $\Delta(1232)$P$_{33}$ resonance, which was treated
differently. According to the phase-shift analyses
of $\pi N$ scattering,
the $\pi N$ amplitude corresponding to
the $P_{33}(1232)$ resonance is elastic
up to $W=1.43~$GeV (see, for example, the
latest GWU analyses \cite{GWU1,GWU2}).
In combination with DR and Watson's theorem, this provides
strict constraints on the multipole amplitudes
$M_{1+}^{3/2}$, $E_{1+}^{3/2}$, $S_{1+}^{3/2}$
that correspond to the $\Delta(1232)$P$_{33}$ resonance.
In particular, as was shown in Ref. \cite{Azn0},
the $W$-dependence of $M_{1+}^{3/2}$
is close to that 
from the GWU analysis \cite{GWU3} at $Q^2=0$
if the same normalizations of the amplitudes
at the resonance position are used.
This constraint on the large 
$M_{1+}^{3/2}$ amplitude plays an important role in the reliable
extraction of the N(1440)P$_{11}$ electroexcitation
amplitudes, because the $\Delta(1232)$P$_{33}$ and N(1440)P$_{11}$ 
states are overlapping.

We have taken into account
all  resonances from the first, second, and third resonance
regions. These are 4- and 3-star resonances
$\Delta(1232)$P$_{33}$, N(1440)P$_{11}$,
N(1520)D$_{13}$, N(1535)S$_{11}$,
$\Delta(1600)$P$_{33}$, $\Delta(1620)$S$_{31}$,
N(1650)S$_{11}$,
N(1675)D$_{15}$,
N(1680)F$_{15}$,
N(1700)D$_{13}$, $\Delta(1700)$D$_{33}$,
N(1710)P$_{11}$, and
N(1720)P$_{13}$.
For the masses, widths, and  $\pi N$ branching
ratios  of these resonances, we used
the mean values of the data presented in the Review of Particle
Physics (RPP) \cite{PDG}. 
In particular for the Roper resonance,
the values $M=1.44~$GeV, $\Gamma=0.35~$GeV,
and $\beta_{\pi N}=0.6$ were taken. Resonances of the fourth
resonance region practically have no influence
in the energy region under investigation
and were not included.

For the values of $Q^2$ under consideration,
the available $ep\rightarrow ep\pi^0$
data are related mostly to the $\Delta(1232)$P$_{33}$
resonance region \cite{Frolov,Joo1,Ungaro}.  
The DESY data \cite{Haidan} at higher energies 
$W=1.14-1.72~$GeV ($Q^2\approx 3~$GeV$^2$) have very 
limited angular coverage. Our analysis showed that
the combined $ep\rightarrow ep\pi^0$
\cite{Frolov,Joo1,Ungaro,Haidan}
and $\vec{e}p\rightarrow en\pi^+$
\cite{Park} data give results that are very
close to those obtained from 
the $\vec{e}p\rightarrow en\pi^+$
data \cite{Park} alone.  
For this reason, and also to avoid mixing data sets with different
systematic uncertainties, in this letter we present 
the results for N(1440)P$_{11}$ obtained from
the analysis of the $\vec{e}p\rightarrow en\pi^+$
data \cite{Park} only. 

At each $Q^2$ available for $\vec{e}p\rightarrow en\pi^+$ 
\cite{Park}, $Q^2= 1.72,~2.05,~2.44,~2.91,~3.48,~4.16~$GeV$^2$, 
we performed two kinds of fits
in both approaches:
(i) The magnitudes of the helicity amplitudes
corresponding to all resonances
listed above were fitted.
(ii) The transverse amplitudes for
the members of the multiplet $[70,1^-]$: $\Delta(1620)$S$_{31}$,
N(1650)S$_{11}$,
N(1675)D$_{15}$, N(1700)D$_{13}$, and $\Delta(1700)$D$_{33}$,
were fixed according to
the single quark transition model  \cite{SQTM},
which relates these amplitudes to those for
N(1520)D$_{13}$ and N(1535)S$_{11}$;
the longitudinal amplitudes of these resonances
and the amplitudes of the resonances
$\Delta(1600)$P$_{33}$ and
N(1710)P$_{11}$, which have small photocouplings \cite{GWU3,PDG}
and are not seen in low $Q^2$ $\pi$ and 2$\pi$ 
electroproduction \cite{Azn065}, 
were assumed to be zero.
The results obtained
for $\Delta(1232)$P$_{33}$, N(1440)P$_{11}$,
N(1520)D$_{13}$, and N(1535)S$_{11}$ in the two fits
were very close to each other. The amplitudes
of the Roper resonance
presented below are the average values of the results
obtained in these fits. The uncertainties arising from 
the averaging procedure we will refer to as uncertainties (I). 
They were included in quadrature into 
the total systematic uncertainties.

The background of both approaches contains Born terms
corresponding to
the $s$- and $u$- channel nucleon exchanges
and $t$-channel pion contribution,
and depends, therefore, 
on the proton, neutron,
and pion form factors.
The background of the UIM contains also 
the $\rho$ and $\omega$ $t$-channel exchanges \cite{Drechsel}
and, therefore, the contribution
of the form factors
$G_{\rho(\omega)\rightarrow\pi\gamma}(Q^2)$.
All of these form factors, except the neutron
electric and $G_{\rho(\omega)\rightarrow\pi\gamma}(Q^2)$ ones,
are known in the region of $Q^2$
under investigation from existing experimental data.
For the proton form factors we used the parameterizations
found for the existing data 
in Ref. \cite{21}. The neutron magnetic form factor
and the pion form factor
were taken from Refs. \cite{22} and \cite{23,24,25,26},
respectively. 
The neutron electric form factor, $G_{E_n}(Q^2)$, is measured
up to $Q^2=1.45~$GeV$^2$ \cite{27}, and Ref. \cite{27}
presents a parameterization 
for all existing data on  $G_{E_n}(Q^2)$ that we used
for extrapolation of
$G_{E_n}(Q^2)$  to $1.7<Q^2<4.2~$GeV$^2$.  
In our final results we accounted for a
systematic uncertainty assuming a
$50\%$ deviation from this parameterization.
There are no measurements of the form factors
$G_{\rho(\omega)\rightarrow\pi\gamma}(Q^2)$;
however, investigations made using both QCD sum rules \cite{28}
and quark model \cite{29} predict a $Q^2$ dependence
of $G_{\rho(\omega)\rightarrow\pi\gamma}(Q^2)$
close to the dipole form factor
$G_d(Q^2)
=1/(1+\frac{Q^2}{0.71GeV^2})^2$.
In our analysis
we assumed that
$G_{\rho(\omega)\rightarrow\pi\gamma}(Q^2)= G_d(Q^2)$,
and introduced in our final results a systematic uncertainty   
that can arise from
a $50\%$ deviation from this assumption.
All of these uncertainties, including those that arise
from the measured proton, neutron, and pion form factors,
were added in quadrature and will be referred to as
systematic uncertainties (II) in our final results.

\begin{figure}[b]
\includegraphics[width=8.7cm, bb=36 414 510 515]{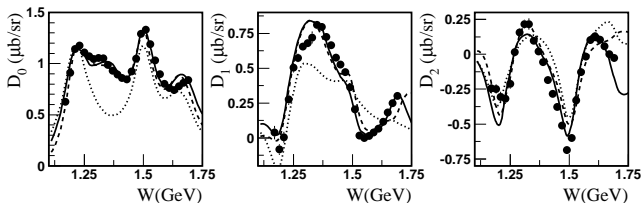}
\caption{
Experimental data for the 3 lowest Legendre moments
of the structure function $\sigma_T+\epsilon \sigma_L$ at
$Q^2=2.05~$GeV$^2$ \cite{Park} in comparison with our
results. The solid and dashed 
curves correspond to the
DR and UIM results, respectively.
The dotted curves are obtained by
switching off the N(1440)P$_{11}$ resonance
in the final DR results. 
}
\label{fig:fig1}
\end{figure}
In Fig. 1, we present the comparison of our results
with the experimental data
for the lowest Legendre moments  of
the structure function
$\sigma_T+\epsilon \sigma_L$
at $Q^2=2.05~$GeV$^2$ \cite{Park}.
The  Legendre moment $D_0^{T+\epsilon L}$ is the
$\cos\theta^*_{\pi}$ independent part of 
$\sigma_T+\epsilon \sigma_L$; it does not
contain interference of different multipole amplitudes and
is related
to the sum of  squares of these amplitudes.
The resonance behavior of the multipole amplitudes is revealed
in $D_0^{T+\epsilon L}$
in the form of enhancements. Resonance structures
related to the resonances
$\Delta(1232)$P$_{33}$, N(1520)D$_{13}$, and N(1535)S$_{11}$
are clearly  seen in
$D_0^{T+\epsilon L}$.
There is a shoulder between the $\Delta$
and $1.5~$GeV peaks, which 
is related to the broad Roper resonance.
To demonstrate this, we present in Fig. 1
the curves obtained by
switching off the N(1440)P$_{11}$ resonance
in the final DR results. 
A fit to the data with the Roper amplitudes put to zero
results in $\chi^2\approx 7$ and gives the dip
in $D_0^{T+\epsilon L}$ of the same size as in Fig. 1.
This clearly shows that the data can not be explained
without the Roper resonance.

To stress the advantage of the investigation of
the Roper resonance in the reaction $\gamma^* p \rightarrow \pi^+ n$,
we note that for this reaction  
the relative contribution
of N(1440)P$_{11}$ in comparison with $\Delta(1232)$P$_{33}$
in $D_0^{T+\epsilon L}$
is four times larger than for $\gamma^* p \rightarrow \pi^0 p$,
because isospin 
$I=\frac{1}{2}$ and $\frac{3}{2}$ resonances enter the
$ep\rightarrow eN\pi$ amplitudes with the coefficients
$\sqrt{\frac{2}{3}},~\sqrt{\frac{1}{3}}$
for $n\pi^+$ in the final state
and $-\sqrt{\frac{1}{3}},~\sqrt{\frac{2}{3}}$ for $p\pi^0$.
                                                                                
The role of N(1440)P$_{11}$  is 
also seen in other  Legendre moments.
In $D_1^{T+\epsilon L}$, the large effect
caused by
switching off this resonance
is connected  with
the interference of $M_{1-}$
corresponding to N(1440)P$_{11}$ with
the non-resonant and N(1535)S$_{11}$ contributions to $E_{0+}$,
which creates a linear dependence of 
$\sigma_T+\epsilon \sigma_L$ in
$\cos\theta^*_{\pi}$.
Due to interference effects like those mentioned above
and to the large width of this state,
the N(1440)P$_{11}$ plays
a significant role
in the entire $W$ range covered by the data.

\begin{table}[t]
\begin{tabular}{|c|c|c|c|c|c}
\hline
$Q^2$&$A_{1/2}~~~~~~~~~~~~~~S_{1/2}$&$N_{data}$&
$\chi^2/N_{data}$\\
(GeV$^2$)&$(10^{-3}$GeV$^{-1/2})$&&\\
\hline
&DR&&\\
\hline
1.72&$72.5\pm 1.0\pm 4.3~~~~24.8\pm 1.4\pm 5.3$&5101&3.1\\
\hline
2.05&$72.0\pm 0.9\pm 4.2~~~~21.0\pm 1.7\pm 5.0$&5844&2.4\\
\hline
2.44&$50.0\pm 1.0\pm 3.2~~~~~9.3\pm 1.3\pm 4.1$&6177&2.1\\
\hline
2.91&$37.5\pm 1.1\pm 2.8~~~~~9.8\pm 2.0\pm 2.3$&6251&2.0\\
\hline
3.48&$29.6\pm 0.8\pm 2.7~~~~~4.2\pm 2.5\pm 2.3$&6105&1.5\\
\hline
4.16&$19.3\pm 2.0\pm 3.9~~~~10.8\pm 2.8\pm 4.5$&5778&1.1\\
\hline
&UIM&&\\
\hline
1.72&$58.5\pm 1.1\pm 4.2~~~~26.9\pm 1.3\pm 5.3$&5101&3.5\\
\hline
2.05&$62.9\pm 0.9\pm 3.3~~~~15.5\pm 1.5\pm 4.9$&5844&2.3\\
\hline
2.44&$56.2\pm 0.9\pm 3.2~~~~11.8\pm 1.4\pm 4.1$&6177&2.1\\
\hline
2.91&$42.5\pm 1.1\pm 2.8~~~~13.8\pm 2.1\pm 2.3$&6251&2.2\\
\hline
3.48&$32.6\pm 0.9\pm 2.6~~~~14.1\pm 2.4\pm 2.0$&6105&1.6\\
\hline
4.16&$23.1\pm 2.2\pm 4.8~~~~17.5\pm 2.6\pm 5.5$&5778&1.1\\
\hline
\end{tabular}
\caption{\label{tab1}The
$\gamma^* p \rightarrow ~$N(1440)P$_{11}$ 
helicity amplitudes
found from the analysis
of  $\pi^+$ electroproduction
data \cite{Park} using DR and UIM.
The first and second uncertainties are, respectively,
the statistical uncertainty from the fit 
and the systematic uncertainties (I) and (II) added in quadrature.
The number of data points, $N_{data}$, and the $\chi^2$ value
per data point are also presented. }
\end{table}
\begin{figure}[t]
\includegraphics[width=8.6cm, bb=32 413 521 647]{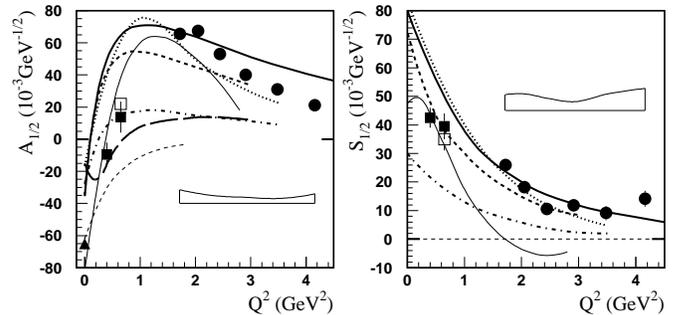}
\caption{
Helicity amplitudes
for the  $\gamma^* p \rightarrow~$N(1440)P$_{11}$ transition.
The full circles are 
our results obtained from
the analysis
of  $\pi^+$ electroproduction
data \cite{Park}.
The bands present the systematic uncertainties
(I,II,III)
added in quadrature; see text.
The full boxes are  the results obtained
from CLAS data \cite{Azn04,Joo1,Joo2,Joo3,Egiyan}; 
open boxes present the results of the combined 
analysis of CLAS
single $\pi$ and $2\pi$ electroproduction
data \cite{Azn065}.
The full triangle at $Q^2=0$ is the RPP estimate \cite{PDG}.
The thick curves correspond to
the light-front relativistic quark models:  dotted, dashed,
dash-dotted, long-dashed, and solid curves are
from Refs. \cite{Capstick1,Weber,Simula,Riska,Quark},  respectively.
The thin solid curves are the predictions obtained for the
Roper resonance treated as
a quark core dressed by a meson cloud \cite{Cano1,Cano2}.
The thin dashed curves are
obtained assuming that N(1440)P$_{11}$
is a q$^3$G hybrid state \cite{Li2}.
}
\label{fig:fig2}
\end{figure}

We now discuss the results for the
$\gamma^* p \rightarrow ~$N(1440)P$_{11}$
helicity amplitudes 
presented in Table \ref{tab1} and Fig. 2.                                                                              
The results obtained using DR and UIM
are given in Table \ref{tab1} separately;
it can be seen that they
are close to each other.
As the non-resonant background of these approaches is
built in conceptually different ways,
we conclude that the model uncertainties
of the obtained results are relatively small.
In Fig. 2 we present average values of the results
obtained within the DR and UIM approaches.
The uncertainties that originate from this averaging
procedure are referred to as 
systematic uncertainties (III) in our final results.
                                                                               
Combined with the information obtained from
the previous CLAS data at $Q^2=0.4,~0.65~$GeV$^2$ 
\cite{Azn04,Azn065,Joo1,Joo2,Joo3,Egiyan},
and that at $Q^2=0$ \cite{PDG}, our results show
the following behavior
of the transverse helicity
amplitude $A_{1/2}$:
being large and negative at $Q^2=0$,
it crosses zero between $Q^2=0.4$ and $0.65~$GeV$^2$
and becomes large and positive
at $Q^2\simeq 2~$GeV$^2$.
With
increasing $Q^2$, this amplitude drops smoothly in magnitude.
The longitudinal helicity amplitude $S_{1/2}$, which is large and
positive at small $Q^2$, drops smoothly with increasing $Q^2$.
                                                                                
In Fig. 2, we compare our results
with model predictions.
These are (i) quark model predictions
\cite{Capstick1,Weber,Simula,Riska,Quark}
where the N(1440)P$_{11}$ is described
as the first radial excitation of the 3$q$ ground state;
(ii) those assuming the
N(1440)P$_{11}$ is
a hybrid state  \cite{Li2}; and (iii)
the results for the
Roper resonance treated as
a quark core (which is a radial excitation of the 3$q$ ground state)
dressed by a meson cloud \cite{Cano1,Cano2}.
                                                                               
It is known that with increasing $Q^2$,
when the momentum transfer becomes larger
than the masses of the constituent quarks, a relativistic
treatment of the electroexcitation of the nucleon resonances,
which is important already at $Q^2=0$,
becomes crucial.
A consistent way to perform the relativistic
treatment of the
$\gamma^* N \rightarrow N^*$  transitions
is to consider
them in light-front (LF) dynamics.
In Fig. 2 we present the results
obtained in the LF quark models 
\cite{Capstick1,Weber,Simula,Riska,Quark}.
All LF approaches \cite{Capstick1,Weber,Simula,Riska,Quark}
give a good description of nucleon form factors, however,
the predictions for the $\gamma^* N \rightarrow~$N(1440)P$_{11}$
helicity amplitudes differ significantly.
This is caused by the large sensitivity
of these amplitudes to
the N and N(1440)P$_{11}$ wave functions \cite{Quark}.                                                                               
The approaches \cite{Capstick1,Weber,Simula,Riska,Quark}
fail to describe the value of the
transverse amplitude $A_{1/2}$ at $Q^2=0$.
This can be an indication of a large meson cloud
contribution to $\gamma^* p \rightarrow~$N(1440)P$_{11}$,
which is expected to be significant at small $Q^2$.
As a confirmation of this assumption, one can consider
the results of Refs. \cite{Cano1,Cano2} where
this contribution
is taken into account,
and a good description of the helicity amplitudes
is obtained at small $Q^2$.
                                                                               
In spite of the
differences, all LF predictions
for the $\gamma^* p \rightarrow~$N(1440)P$_{11}$
helicity amplitudes
have common features
that agree
with the results extracted from the experimental data:
(i) the sign of the transverse amplitude
$A_{1/2}$ at $Q^2=0$
is negative,
(ii) the sign
of the longitudinal
amplitude $S_{1/2}$ is positive,
(iii) all LF approaches predict the sign change
of the transverse amplitude $A_{1/2}$ at small $Q^2$.
We take this qualitative agreement as evidence in favor
of the N(1440)P$_{11}$ resonance as a radial excitation
of the 3$q$ ground state. Final confirmation of this conclusion requires
a complete simultaneous description of the nucleon
form factors and the  $\gamma^* p \rightarrow~$N(1440)P$_{11}$ 
amplitudes.
This will allow us to  find the magnitude of the meson cloud 
contribution, and
to better specify the $N$ and N(1440)P$_{11}$ wave functions.
To achieve a satisfactory description at large $Q^2$, it may be
necessary to take into account quark form factors,
as well as other
effects, such as the quark mass dependence
on the momentum transfer.
                          
The results of Refs. \cite{Li1,Li2}, where
N(1440)P$_{11}$ is treated as
a hybrid state,  are obtained via non-relativistic calculations.
Nevertheless the suppression
of the longitudinal amplitude $S_{1/2}$ has its physical
origin
in the fact that the longitudinal
transition operator for the vertex $\gamma q\rightarrow q G$
requires both a spin and angular momentum flip by one unit,
while the angular momenta of quarks in the N and N(1440)P$_{11}
\equiv $q$^3$G are equal to 0.
This makes this result practically independent
of relativistic effects.
The predicted suppression of the longitudinal
amplitude $S_{1/2}$
strongly disagrees with the experimental results.
                                                                                           
In summary, for the first time the transverse and longitudinal helicity
amplitudes of the  $\gamma^* p \rightarrow~$N(1440)P$_{11}$ transition
are extracted from experimental data
at high $Q^2$. The results are obtained from differential
cross sections and longitudinally polarized
beam asymmetries for $\pi^+$ electroproduction on protons
at $W=1.15-1.69~$GeV \cite{Park}.
The data were analyzed using two
conceptually different approaches, DR and UIM,
which give close results.

Comparison with quark model predictions provides strong 
evidence in favor of
N(1440)P$_{11}$ as a first radial excitation of the 3$q$ ground state. 

The results for the longitudinal helicity amplitude
confirm our conclusion made from the 
previous analysis of CLAS $\vec{e}p\rightarrow ep\pi^0,en\pi^+$
data for  $Q^2< 1~$GeV$^2$ \cite{Azn04} that 
the presentation of
the Roper resonance as a q$^3$G hybrid state
is ruled out.

This work was supported
in part by the U.S. Department of Energy and the National
Science Foundation, the Korea Research Foundation,
the French Commissariat a l'Energie Atomique, and the
Italian Instituto Nazionale di Fisica Nucleare.
Jefferson Science Associates, LLC, operates Jefferson Lab
under U.S. DOE contract DE-AC05-060R23177.


\begin{thebibliography}{999} 
\bibitem{Capstick1} S. Capstick and B. D.
Keister, Phys. Rev. D $\bf{51}$, 3598 (1995).
                                                                                
\bibitem{Capstick2} S. Capstick and N. Isgur,
Phys. Rev. D $\bf{34}$, 2809 (1986).
                                                                                
\bibitem{Richard} J.-M. Richard,
Phys. Rep. $\bf{212}$, 1 (1992).
                                                                                
\bibitem{Glozman} L. Ya. Glozman and D. O. Riska,
Phys. Rep. $\bf{268}$, 263 (1996).

\bibitem{Li1}   Z. P. Li,
Phys. Rev. D $\bf{44}$, 2841 (1991).
                                                                                
\bibitem{Li2}   Z. P. Li, V. Burkert, and Zh. Li,
Phys. Rev. D $\bf{46}$, 70 (1992).
                                                                                
\bibitem{Cano1}  F. Cano, P. Gonz\'alez,
S. Noguera, and B. Desplanques, Nucl. Phys. A $\bf{603}$, 257 (1996).
                                                                                
\bibitem{Cano2}  F. Cano and P. Gonz\'alez,
Phys. Lett. B $\bf{431}$, 270 (1998).

\bibitem{Krewald}  O. Kreil, C. Hanhart, S. Krewald, and J. Speth,
Phys. Rev. C $\bf{62}$, 025207 (2000).

\bibitem{Dillig}  M. Dillig and J. Schott,
Phys. Rev. C $\bf{75}$, 067001 (2007).

\bibitem{Park} K. Park et al., CLAS Collaboration,
Phys. Rev. C $\bf{77}$, 015208 (2008).

\bibitem{Azn0} I. G. Aznauryan,
Phys. Rev. C $\bf{67}$, 015209 (2003).

\bibitem{Azn04} I. G. Aznauryan, V. D. Burkert, H. Egiyan,
et al., Phys. Rev. C $\bf{71}$, 015201 (2005).

\bibitem{Azn065} I. G. Aznauryan, V. D. Burkert,
et al., Phys. Rev. C $\bf{72}$, 045201 (2005).

\bibitem{Drechsel} D. Drechsel, O. Hanstein, S. Kamalov,  and L. Tiator,
Nucl. Phys. A $\bf{645}$, 145 (1999).

\bibitem{GWU1} R. A. Arndt, W. J. Briscoe, I. I. Strakovsky,
and R. L. Workman,
Phys. Rev. C $\bf{69}$, 035213 (2004).

\bibitem{GWU2} R. A. Arndt, W. J. Briscoe, I. I. Strakovsky,
and R. L. Workman,
Phys. Rev. C $\bf{74}$, 045205 (2006).

\bibitem{GWU3} R. A. Arndt, W. J. Briscoe, I. I. Strakovsky,
and R. L. Workman,
Phys. Rev. C $\bf{66}$, 055213 (2002).

\bibitem{PDG} W.-M. Yao et al. [Particle Data Group],
Journal of Physics G  $\bf{33}$, 1 (2006).

\bibitem{Frolov} V. V. Frolov et al., 
Phys. Rev. Lett. $\bf{82}$, 45 
(1999).

\bibitem{Joo1} K. Joo et al., 
CLAS Collaboration, Phys. Rev. Lett. $\bf{88}$, 122001 
(2002).

\bibitem{Ungaro} M. Ungaro et al., 
CLAS Collaboration, Phys. Rev. Lett. $\bf{97}$, 112003 
(2006).

\bibitem{Haidan} R. Haidan, PhD thesis, University of Hamburg,
Hamburg, 1979. 

\bibitem{SQTM} V. D. Burkert et al.,
Phys. Rev. C $\bf{67}$, 035204 (2003).

\bibitem{21} J. Arrington, W. Melnitchouk,
and J. A. Tjon,  
Phys. Rev. C $\bf{76}$, 035205 (2007).

\bibitem{22} W. K. Brooks et al., Nucl. Phys. A $\bf{755}$, 261 
(2005).

\bibitem{23} C. J. Bebek et al., Phys. Rev. D $\bf{13}$, 25 (1976).

\bibitem{24} C. J. Bebek et al., Phys. Rev. D $\bf{17}$, 1693 (1978).

\bibitem{25} T. Horn  et al., Phys. Rev. Lett. $\bf{97}$, 192001 (2006).

\bibitem{26} V. Tadevosyan et al., Phys. Rev. C  $\bf{75}$, 055205 
(2007).

\bibitem{27} R. Madey et al., Phys. Rev. Lett. $\bf{91}$, 122002 (2003).

\bibitem{28} V. Eletski and Ya. Kogan, Yad. Fiz. $\bf{39}$, 138 (1984).

\bibitem{29} I. Aznauryan and K. Oganessyan, Phys. Lett. B $\bf{249}$,
309 (1990).

\bibitem{Joo2} K. Joo et al., 
CLAS Collaboration, Phys. Rev. C $\bf{68}$, 032201 
(2003).

\bibitem{Joo3} K. Joo et al., 
CLAS Collaboration, Phys. Rev. C $\bf{70}$, 042201 
(2004).

\bibitem{Egiyan} H. Egiyan et al., 
CLAS Collaboration, Phys. Rev. C $\bf{73}$, 025204 
(2006).

\bibitem{Weber} H. J. Weber, Phys. Rev.
C $\bf{41}$, 2783 (1990).
                                                                                
\bibitem{Simula} F. Cardarelli, E. Pace, G. Salme, and S. Simula,
Phys. Lett. B $\bf{397}$, 13 (1997).
                                                                                
\bibitem{Riska} B. Juli\'a-D\'iaz,
D. O. Riska, and F. Coester, Phys. Rev.
C $\bf{69}$, 035212 (2004).
         
                                                                               
\bibitem{Quark} I. G. Aznauryan, Phys. Rev.
C $\bf{76}$, 025212 (2007).
                                                                                
\end{thebibliography}
\end{document}